\documentclass[12pt]{article}
\usepackage{amssymb}
\usepackage{amsfonts}
\usepackage{amsmath,amsthm}
\usepackage{graphics,epsfig}
\usepackage{subfigure}
\usepackage{graphicx,epsfig, color }
\oddsidemargin 10mm \numberwithin{equation}{section}
\begin{document}
\begin{center}
{{\bf{Thermodynamic phase transition and Joule Thomson adiabatic
expansion for dS/AdS Bardeen Black Holes with consistent 4D
Gauss-Bonnet gravity }} \vskip 1 cm
H.Ghaffarnejad$^{(a)}$\footnote{E-mail
address:hghafarnejad@semnan.ac.ir},E.Ghasemi$^{(a)}$\footnote{E-mail
address:e\_ghasemi@semnan.ac.ir},
E.Yaraie,$^{(a,b)}$\footnote{E-mail address:eyaraie@semnan.ac.ir}
and M.Farsam $^{(a,b)}$\footnote{E-mail
address:mhdfarsam@semnan.ac.ir}
}\\
\vskip 0.1 cm
{\textit{Faculty of Physics,Semnan University, 35131-19111, Semnan, Iran$^{(a)}$}\\and\\
\textit{Instituut-Lorentz for Theoretical Physics, $\Delta$-ITP,
Leiden University, Niels Bohrweg 2, Leiden 2333 CA,The
Netherlands$^{(b)}$}}
\end{center}
\begin{abstract}
Instead of the work \cite{1} which in according to the Lovelock
theorem it could not applicable for all types of 4D curved
spacetimes of Einstein Gauss Bonnet (EGB) gravity, authors of the
work \cite{2} applied break of diffeomorphism property to present
a consistent EGB gravity theory. In this work we use the latter
model by adding a nonlinear electromagnetic field lagrangian
density to study affects of GB coupling constant into the
thermodynamic phase transition and Joule Thomson (JT) adiabatic
expansion of a 4D $dS/AdS$ GB Bardeen black hole.
\end{abstract}
 \section{Introduction} From the theoretical point view we know the black holes are
made from metric solutions of the Einstein's metric field equation
with no temperature and so they are not supposed to show any
thermodynamic behavior. For the first time, Hawking presented an
important theorem where the event horizon of the black holes
should never be decreased because all objects are absorbed by them
\cite{3}. This is called now as the Hawking's area theorem. After
this presentation Bekenstein suggested that for the black hole
should be assigned an entropy appropriate to the area of its
horizon\cite{4}. In analogy with the thermodynamics rules of the
ordinary systems, four laws proposed for the black holes
thermodynamics. But by considering this analogy, there was
obtained a problem for thermodynamics of the black holes as
follows. Actually the first law of the thermodynamics of the black
holes lacks the pressure and volume components. Because there is
no a clear concept for thermodynamic volume and pressure of the
black hole. The first idea to solve the pressure problem led to
the consideration of a negative cosmological constant
\cite{5,6,7,8,9} which it is called conjugate variable for the
thermodynamic volume. There are done a lot of research where the
pressure-volume (PV) criticality of thermodynamics of AdS black
holes mimic thermodynamic behavior of the well known Van der Waals
ordinary
gases\cite{10,11,12,13,14,15,16,17,18,19,20,21,22,23,24,25,26,27,28}.
Recently Glavan and Lin released a paper \cite{29} in which an
alternative generally covariant gravity theory is defined which in
$4D$ curved space-times can propagates just massless gravitons by
bypassing the Lovelock`s theorem. This alternative higher order
derivatives gravity theory has two correction terms called as
Gauss-Bonnet topological invariant and cosmological constant
respectively. In 4D curved spacetimes  the Gauss-Bonnet coupling
constant parameter diverges to an infinite value. In this singular
limit the Gauss-Bonnet topological invariant term gives rise
non-trivial contributions to the gravitational  dynamics, while
preserving the number of degrees of freedom of graviton and being
free from Ostrogradsky instability. They reported some appealing
corrections to the dispersion relation of cosmological
tensor-scalar modes in the cosmological space times and also
singularity resolution in the spherically symmetric space times.
As an spherically symmetric static black hole metric solutions of
this model, authors of the work \cite{30} obtained a Bardeen type
of the black hole solutions and generated its thermodynamic
variables via the horizon calculation. They obtained a critical
location for the black hole horizon where the corresponding
Hawking temperature raises to a maximum value for which a
second-order phase transition is happened because the heat
capacity diverges to infinity. Existence of these appealing
thermodynamic behavior encourages us to study Joule-Thomson
adiabatic free expansion phenomena for this black hole given in
the AdS background. If we want to describe this expansion briefly,
this is down as follows. In fact this expansion is happened when a
gas is allowed to move from a high pressure region to a low
pressure one without to change its enthalpy. As it established in
the above, the black hole mass would be taken as enthalpy in an
extended thermodynamic phase space, so during the Joule Thomson
expansion phenomena the mass remains constant (isentropic
process). In presence of this expansion, the black hole usually
could reach to one of two the heating or cooling phases finally.
After to present the pioneer works about the black hole Joule
Thomson expansion given by \cite{31,32,33}, many other scientists
investigated this phenomena for several black holes interacting
with many types of the material fields
\cite{34,35,36,37,38,39,40,41,42,43,44,45,46,47,48,49}. To study
this phenomena one can usually investigate the inversion curves
which mimics behavior of the van der Waals fluid. In general there
are several AdS black holes which behave as different with respect
to the van der Waals fluid and they do not mimic completely with
the inversion curves (see for instance \cite{30,31}). In this
work, we want to examine the possibility of the emergence of the
expansion phenomenon of Joule Thomson for $dS/AdS$ GB Bardeen
black hole. In fact we will see importance of parameters of this
black hole namely the magnetic charge and the GB coupling constant
into its heating-cooling phase transition.The paper is organized
as follows.\\ In the second section we define briefly the
consistent 4D GB gravity model given by \cite{2} and as an
application we consider the Bardeen black hole nonlinear
electromagnetic field lagrangian density. In the third section we
derive metric field equations of $dS/AdS$ GB Bardeen black hole
and solve them by using some physical arguments. In fourth section
we calculate the horizon equation, the Hawking temperature, the
equation of state and the Joule Thomson coefficient for different
regimes of the 4D GB dS/AdS black hole mass density function. Then
by plotting P-V diagrams at constant temperature and also the
isenthalpic T-P curves we study possibility of the phase
transitions and JT expansion of the black hole under
consideration.
 Fifth section is dedicated to conclusion and outlook of the work.
\section{Consistent 4D EGB gravity } By
according to the work \cite{2} we define consistent EGB gravity in
$D\to4$ limit with the first term of the lagrangian density in the
following action functional in which the second term of the
lagrangian density $\mathcal{L}_{matter}$ denotes to source part.
\begin{equation}\label{action}I=\frac{1}{16\pi G}\int
dtd^3xN\sqrt{\gamma}\big(\mathcal{L}_{EGB}^{4D}+\mathcal{L}_{matter}\big),
\end{equation} where
\begin{equation}\label{lagEGB}\mathcal{L}_{\mathrm{EGB}}^{4D}=2R-2\Lambda-\mathcal{M}\end{equation}
$$+\frac{\tilde{\alpha}}{2}[8R^2-4R\mathcal{M}-\mathcal{M}^2-\frac{8}{3}(8R_{ij}R^{ij}-4R_{ij}\mathcal{M}^{ij}-\mathcal{M}_{ij}\mathcal{M}^{ij})],
$$ and $G$ is the Newton`s gravitational coupling constant. $R$
and $R_{ij}$ are  the Ricci scalar and the Ricci tensor of the
spatial 3-metric $\gamma_{ij}$ respectively. In the definition
(\ref{lagEGB}) we have
\begin{equation}\mathcal{M}_{ij}=R_{ij}+\mathcal{K}_k^k\mathcal{K}_{ij}-\mathcal{K}_{ik}\mathcal{K}^{k}_j,~~\mathcal{M}=\mathcal{M}_i^i
\end{equation} with
\begin{equation}\mathcal{K}_{ij}=\frac{1}{2N}(\dot{\gamma}_{ij}-2D_iN_j-2D_jN_i-\gamma_{ij}D_kD^k\lambda_{GF}).\end{equation} Here, dot $\dot{~}$ denotes derivative with respect to the
time $t$ and all the effects of the constraint stemming  from  the
$gauge-fixing$ (GF) are  now  encoded  in lagrange multiplier
$\lambda_{GF}$. $D_i$ is spatial covariant  derivative and
re-scaled regular EGB coupling constant $\tilde{\alpha}$ is
defined versus the irregular GB coupling constant $\alpha_{GB}$
such that $\tilde{\alpha}=(D-4)\alpha_{GB}$ which in limits of
$D\to4 $ dimensions become finite.  The above EGB gravity action
functional satisfies the following gauge condition for all
spherically symmetric and cosmological backgrounds (see \cite{2}
and \cite{50}).
\begin{equation}\sqrt{\gamma}D_kD^k(\pi^{ij}\gamma_{ij}/\sqrt{\gamma})\approx0.\end{equation} In fact the above EGB action functional is generated from
ADM decomposition of the 4D background metric as $1+3$ dimensions
such that \begin{equation}\label{met1}ds^2=g_{\mu\nu}dx^\mu
dx^\nu=-N^2dt^2+\gamma_{ij}(dx^i+N^idt)(dx^j+N^jdt)\end{equation}
where $N, N_i, \gamma_{ij}$ are the lapse function, the shift
vector, and the spatial metric respectively. $\gamma$ factor in
the action functional (\ref{action}) is absolute value of
determinant of the spatial 3-metric $\gamma_{ij}.$ This ADM
decomposition is done on the background metric to remove divergent
boundary term of the higher order metric derivatives in the GB
term of the action functional (\ref{action}) in general 4
dimensional form \cite{2}. First term in the  theory  defined by
(\ref{action}) has the time re-parametrization symmetry $t\to
t=t(t^\prime).$ We now set the matter source $I_{\mathrm{matter}}$
to be action of a nonlinear electromagnetic antisymmetric Maxwell
field $F_{\mu\nu}$ with Ayon Beato Garcia form of the lagrangian
density as follows.
\begin{equation}\label{ABG}\mathcal{L}_{matter}=\mathcal{L}_{ABG}(F)=\frac{12M_{ADM}}{Q^3}\bigg(\frac{\sqrt{2Q^2F}}{1+\sqrt{2Q^2F}}\bigg)^\frac{5}{2}
\end{equation} where $F=F_{\mu\nu}F^{\mu\nu},$ $M_{ADM}$ is ADM mass of the
black hole and $Q$ is Bardeen magnetic charge (see eq. (8) in ref.
\cite{51}). \section{4D $dS/AdS$ GB Bardeen Black Holes} By
comparing the line element (\ref{met1}) with general form of a
spherically symmetric static 4D metric field
\begin{equation}\label{line1}ds^2=-e^{2A(r)}\bigg(1-\frac{2M(r)}{r}\bigg)dt^2+\frac{dr^2}{1-\frac{2M(r)}{r}}+r^2d\theta^2+r^2\sin^2\theta
d\varphi^2\end{equation} we infer that the lapse function $N$, and
the shift vector $N_i$ and the spatial metric components
$\gamma_{ij}$ and gauge fixing lagrange multiplier $\lambda_{GF}$
should be $r$ dependent so that we can write
\begin{equation}
N=e^{A(r)}\sqrt{1-\frac{2M(r)}{r}},~~~N_{r,\theta,\varphi}=0,\end{equation}$$\gamma_{rr}=\frac{1}{1-\frac{2M(r)}{r}}~
~~\gamma_{\theta\theta}=r^2,~
~~\gamma_{\varphi\varphi}=r^2\sin^2\theta,~~~\lambda_{GF}=\lambda_{GF}(r).$$
By substituting (\ref{line1}) into (\ref{lagEGB}) we obtain
\begin{equation}\label{EGB}\mathcal{L}^{4D}_{EGB}=R(\gamma_{ij})-2\Lambda+12q^2+\frac{\tilde{\alpha}}{2}\bigg[3R^2(\gamma)+\frac{88}{3}q^2R(\gamma)-272q^4-8R_{ij}(\gamma)R^{ij}(\gamma)\bigg]\end{equation} in
which
\begin{equation}\label{q}q(r)=\frac{e^{-A(r)}}{r^2}\bigg[r^2\lambda^\prime_{GF}(r)\bigg]^\prime,~~~\mathcal{K}_{ij}=-q\gamma_{ij}
\end{equation}\begin{equation}\label{R}R(\gamma)=-\frac{4M^\prime}{r^2},~~~R_{ij}(\gamma)R^{ij}(\gamma)=\frac{6}{r^6}(M-rM^\prime)^2\end{equation}
and $\prime$ denotes derivative with respect to $r$. From the
Maxwell equations, we can prove that the magnetic field has the
form \cite{51}
\begin{equation}F_{\theta\varphi}(\theta)=Q\sin\theta\end{equation}which for the metric equation (\ref{line1}) the EM lagrangian
density
become\begin{equation}F=\frac{1}{4}F_{\mu\nu}F^{\mu\nu}=\frac{Q^2}{2r^4}\end{equation}
for which (\ref{ABG}) leads to the following form \cite{51}.
\begin{equation}\label{radialABG}\mathcal{L}_{Bardeen}(r)=\frac{12M_{ADM}}{Q^3}\bigg(\frac{Q^2}{r^2+Q^2}\bigg)^\frac{5}{2}.\end{equation}
 By adding (\ref{radialABG}) and by substituting  (\ref{EGB})
and (\ref{R}) and by integrating the action functional
(\ref{action}) on the 2-sphere $0\leq\theta\leq\pi$,
$0\leq\varphi\leq2\pi$ we obtain
\begin{equation}\label{acttot}I=\frac{1}{4G}\int dt\int dr
r^2e^{A(r)}\bigg\{
-\frac{4M^\prime}{r^2}-2\Lambda+12q^2+\frac{12M_{ADM}}{Q^3}\bigg(\frac{Q^2}{r^2+Q^2}\bigg)^\frac{5}{2}\end{equation}$$
-\tilde{\alpha}\bigg[\frac{176M^{\prime}q^2}{3r^2}+136q^4+\frac{24M^2}{r^6}-\frac{48MM^\prime}{r^5}\bigg]\bigg\}.$$
Euler Lagrange equation for $q$ reads
\begin{equation}q\bigg[12-\tilde{\alpha}\bigg(\frac{176M^\prime}{3r^2}+136q^2\bigg)\bigg]=0\end{equation} which has two different solutions
as \begin{equation}\label{qq}
q_1=0,~~~q_2=\frac{\pm1}{\sqrt{136}}\sqrt{\frac{12}{\tilde{\alpha}}-
\frac{176M^\prime}{3r^2}}.\end{equation} By substituting these two
different gauge fixing conditions into the equation (\ref{q}) and
the action functional (\ref{acttot}) we obtain
\begin{equation}\label{lambda}\lambda^{(1)}_{GF}(r)\sim\frac{1}{r}\end{equation}
\begin{equation}\lambda_{GF}^{(2)}(r)=\int^r\frac{dr^\prime}{r^{\prime2}}\int^{r^\prime} {r^{\prime\prime}}^2q_2(r^{\prime\prime})
e^{A(r^{\prime\prime})}dr^{\prime\prime} \end{equation} and
\begin{equation}\label{acttot1}I_1=I_2=\frac{1}{4G}\int dt\int dr r^2
e^{A(r)}\bigg\{-\frac{4M^\prime}{r^2}-2\Lambda+
\frac{12M_{ADM}}{Q^3}\bigg(\frac{Q^2}{r^2+Q^2}\bigg)^\frac{5}{2}\end{equation}$$
+\tilde{\alpha}\bigg[-\frac{24M^2}{r^6}+\frac{48MM^\prime}{r^5}\bigg]\bigg\}.$$
This shows that two different gauge fixing conditions $q_{1,2}$
reach to similar action functional $I_1=I_2$ and so similar metric
solutions. The Euler Lagrange equations for the function $A(r)$
and the mass distribution function $M(r)$ reduce to the following
relations respectively. \begin{equation}\label{N11}
\frac{4M^\prime}{r^2}=\frac{\frac{12M_{ADM}}{Q^3}\big(\frac{Q^2}{r^2+Q^2}\big)^\frac{5}{2}-
2\Lambda-\frac{24\tilde{\alpha}M^2(r)}{r^6}}{1-\frac{12\tilde{\alpha}M(r)}{r^3}}\end{equation}
and
\begin{equation}\label{M11}A^\prime(r)=\frac{\frac{-24\tilde{\alpha}M(r)}{r^4}}{1-\frac{12\tilde{\alpha}M(r)}{r^3}}.\end{equation}
 Now we are in position to solve the above nonlinear differential equations. Here we try to obtain
 analytical solutions for the metric equations which are applicable in studying of the black hole thermodynamic.
  To do so we  pay attention to the equation
 (\ref{N11}) which up to the Gauss Bonnet term $\tilde{\alpha}=0$ reads to the following
 solution.
\begin{equation}\label{m0}M_0(r)=\frac{M_{ADM}r^3}{(r^2+Q^2)^\frac{3}{2}}-\frac{\Lambda r^3}{6},\end{equation} for which \begin{equation}A_0=constant=0.\end{equation}
By substituting $\Lambda=0$ into the above zero order solution of
the mass function we obtain $\lim_{r\to\infty}M_0(r)=M_{ADM}$ and
the metric equation (\ref{line1}) reads to the original form of
the Bardeen black hole. By applying (\ref{m0}) one can show that
the metric fields reaches to a vacuum (anti) de Sitter metric
solution asymptotically at center of the black hole and
Schwarzschild (anti) de Sitter form of a black hole asymptotically
at far from its central regions such that
\begin{equation}\label{less}\lim_{r<<|Q|}\bigg(1-\frac{2M_0(r)}{r}\bigg)\sim\bigg(1-\frac{\bar{\Lambda}}{3}r^2\bigg),~~~\bar{\Lambda}=\frac{6M_{ADM}}{Q^3}-\Lambda,\end{equation}
and \begin{equation}
\lim_{r>>|Q|}\bigg(1-\frac{2M_0(r)}{r}\bigg)\sim\bigg(1-\frac{2M_{ADM}}{r}
+\frac{\Lambda}{3}r^2\bigg).\end{equation} By looking at the
asymptotically solution (\ref{less}) one can infer that the
quantity $\frac{6M_{ADM}}{Q^3}$ behaves as alternative
cosmological constant to make as nonsingular the central region of
the black hole.\\
However we try to solve the equation (\ref{N11}) to obtain an
analytic solution for the mass function $M(r)$ as follows.
According to the zero order solution (\ref{m0}), it is easy to
write the equation (\ref{N11}) as follows.
\begin{equation}\frac{4M^\prime}{r^2}=\frac{\frac{4M_0^\prime}{r^2}-\frac{24\tilde{\alpha}M^2}{r^6}}{1-\frac{12\tilde{\alpha}M}{r^3}}
\end{equation}
which is equivalent with
\begin{equation}\label{eqm}r^2\frac{d}{dr}(M-M_0)=\frac{6\tilde{\alpha}[r(M^2)^\prime-M^2]}{r^2}=6\tilde{\alpha}\frac{d}{dr}\bigg(\frac{M^2}{r}\bigg)
.\end{equation} This equation is a nonlinear differential equation
for mass function $M(r)$ and its nonlinearity is generated from
the GB parameter $\tilde{\alpha}.$ To study thermodynamics of this
black hole metric we need an analytic solution which for nonlinear
differential equations it is possible to obtain via perturbation
series method for small $\tilde{\alpha}.$ To do so we define
\begin{equation}M(r;\tilde{\alpha})=M_0(r)+\tilde{\alpha}M_1(r)+\tilde{\alpha}^2M_2(r)+O(3)\end{equation} in which we defined
\begin{equation}M_0(r)=M(r;0),~~~M_1(r)=\frac{\partial M}{\partial \tilde{\alpha}}\bigg|_{\tilde{\alpha}=0},~~~M_2(r)=\frac{\partial^2 M}{\partial \tilde{\alpha}^2}\bigg|_{\tilde{\alpha}=0}
\end{equation} and so on. By substituting the above series expansion into the equation (\ref{eqm}) and by keeping coefficients of the
GB parameter
 as different linear differential equations and solving them as step by step we can obtain for up to third order term
 \begin{equation}M_1(r)=\int\frac{6}{r^2}d\bigg(\frac{M_0^2(r)}{r}\bigg)=\frac{6M_0^2(r)}{r^3}+12\int\frac{M_0^2(r)}{r^4}dr\end{equation}
and
\begin{equation}M_2(r)=\int\frac{12}{r^2}d\bigg(\frac{M_0M_1}{r}\bigg)=\frac{12M_0M_1}{r^3}+\int\frac{24M_0M_1}{r^3}dr.\end{equation}
For small value of the GB parameter $\tilde{\alpha}$ the first
order solution of the above mass function is enough to seek how
can affects the GB parameter on the thermodynamics properties of
this kind of black holes. Linear order solution of the mass
function reads
\begin{equation}\label{malpha}M(r)\approx M_0(r)+\tilde{\alpha}M_1(r)\end{equation}
for which $M_0(r)$ should be substituted via the equation
(\ref{m0}) and $M_1(r)$ is calculated trivially as follows.
\begin{equation}M_1(r)=\frac{6M_0^2(r)}{r^3}+\frac{\Lambda^2 r^3}{9}+\frac{4\Lambda M_{ADM}r}{\sqrt{r^2+Q^2}}+
\end{equation}
$$\frac{3M^2_{ADM}}{2Q^3}\arctan\bigg(\frac{r}{Q}
\bigg)-4\Lambda M_{ADM}\arcsin h\bigg(\frac{r}{Q}
\bigg)+\frac{3M^2_{ADM}r}{2Q^2(r^2+Q^2)}-\frac{3M^2_{ADM}r}{(r^2+Q^2)^2}$$
 for
which the equation (\ref{M11}) reads
\begin{equation}\label{A}A(r)=4\tilde{\alpha}\bigg\{\Lambda\ln\bigg(\frac{r}{Q}\bigg)-\frac{6M_{ADM}}{Q^2\sqrt{Q^2+r^2}}+\frac{6M_{ADM}}{Q^3}\arctan h \bigg(\frac{Q}{\sqrt{Q^2+r^2}}
\bigg)\bigg\}.\end{equation}  For this obtained solutions the
metric components are
\begin{equation}\label{grr}g^{rr}(r)=1-\frac{2(M_0(r)+\tilde{\alpha}M_1(r))}{r},~~~g_{tt}(r)=-e^{2A(r)}g_{rr}(r).\end{equation}In the next section we
study thermodynamic behavior of 4D GB Bardeen black hole horizon
with this mass distribution function.
\section{Thermodynamics of 4D GB dS/AdS Bardeen black hole} For the mass distribution solution (\ref{malpha}) the exterior horizon
equation of the 4D GB dS/AdS Bardeen black hole is obtained by
solving the equation $g^{rr}(r_+)=0$ given by (\ref{grr}) as
follows.
\begin{equation}\label{hor}1=x^2\bigg(\frac{2m}{(x^2+1)^\frac{3}{2}}-\frac{\lambda}{3}\bigg)+\mu\bigg\{\frac{3m^2(x^4+4x^2-1)}{(x^2+1)^3}
+\frac{5x^2\lambda^2}{9}\end{equation}
$$+\frac{8m\lambda}{(x^2+1)^\frac{3}{2}}+3m^2\frac{\arctan
{x}}{x}-8m\lambda\frac{\arcsin h {x}}{x}\bigg\}$$ in which we
defined dimensionless ADM mass $m$ and dimensionless cosmological
parameter $\lambda,$ dimensionless horizon position $x$ and
dimensionless GB parameter $\mu$ as follows.
\begin{equation}\label{defff}m=\frac{M_{ADM}}{Q},~~~\lambda=\Lambda Q^2,~~~x=\frac{r_+}{Q},~~~\mu=\frac{\tilde{\alpha}}{Q^2}.\end{equation}
The Hawking temperature of this kind of the black hole is obtained
vs the surface gravity such that
\begin{equation}\label{TT}T=\frac{-g_{tt}^\prime}{4\pi}\bigg|_{r_+}
=\frac{e^{2A(r_+)}}{4\pi}\bigg(\frac{2M(r_+)}{r_+^2}-\frac{2M^\prime(r_+)}{r_+}\bigg)
\end{equation} which by substituting $r_+=2M(r_+),$
the equation $M^\prime(r)$ given by (\ref{N11}) and the solution
(\ref{A}) reads to the following dimensionless form.
\begin{equation}\label{tx}t(x)=4\pi QT=\frac{x^{1+\lambda\mu}}{(x^2-6\mu)}\bigg[1-\frac{3\mu}{x^2}-\frac{6m x^2}{(x^2+1)^\frac{5}{2}}+\lambda x^2\bigg]
\end{equation}
$$\times\exp\bigg\{6\mu m\bigg[\arctan
h\bigg(\frac{1}{\sqrt{x^2+1}}\bigg)-\frac{1}{\sqrt{x^2+1}}\bigg]\bigg\}$$
where we substituted the dimensionless quantities (\ref{defff}).
To study thermodynamic behavior of this kind of black hole it is
simpler to use asymptotic behavior of the event horizon equation
and the Hawking temperature equation instead of their exact forms.
In fact calculations with the exact equations take on complicated
forms. To do so we can study two different approaches as $x<1$ and
$x>1.$ In the case $x>1$ asymptotic series forms of the horizon
equation and the Hawking temperature take on divergent forms and
so give not suitable physical situations for the p-v curves at
constant temperature and other thermodynamic diagrams but for
small scale black holes $x<1$ the behavior of p-v diagram predicts
small to large black hole phase transition and vice versa
dependent to
 choose numeric values which we use for the GB parameter $\mu$.\\
 By according to the GB parameter $\mu$
which takes on some small values $\mu<1$ let us we start to
substitute the mass parameter $m(x)$ given by the horizon equation
(\ref{hor}) into the temperature equation (\ref{tx}) which takes
on a long length form and so we calculate its series expansion
about small values of the parameters $\mu<1$ and $x<1$ such that
\begin{equation}\label{tap}t\approx\frac{(213\mu-20)}{10x}-\frac{6\mu}{x^3}\bigg(4+\ln\bigg|\frac{2}{x}\bigg|\bigg)+\frac{2\mu(4-\ln2)\lambda}{x}+O(x,x\ln x,\mu^2)
\end{equation}
and
\begin{equation}m\approx\frac{5\mu}{2x^4}+\frac{A}{x^2}+B+O(x^2,\mu^2)\end{equation}
where we defined
\begin{equation}A=\frac{1}{2}+\frac{99}{20}\mu-\lambda\mu,~~~B=\frac{3}{4}+\frac{\lambda}{6}+\frac{1629}{560}\mu
-\frac{\lambda^2\mu}{3}-\frac{11\lambda\mu}{10}.\end{equation} In
the above series expansion forms the $O$ terms are not divergent
for small values of $x<1$ and $\mu<1$ and so we can remove them
when we study thermodynamic behavior of small scale black holes.
To obtain equation of state we set thermodynamic specific volume
$v$ and dS/AdS pressures respectively as follows.
\begin{equation}\label{vA} AdS:~~~\lambda=-8\pi p,~~~v=\frac{-16\pi\mu(4-\ln2)}{x}\end{equation} and
\begin{equation}\label{vd}dS:~~~\lambda=8\pi p,~~~v=\frac{16\pi\mu(4-\ln2)}{x}\end{equation} for which the temperature equation (\ref{tap}) reads
 \begin{equation}\label{ttt}dS/AdS:~~~t=pv+6\mu\epsilon\bigg(
\frac{v}{16\pi\mu(4-\ln2)}\bigg)^3\bigg(4+\ln\bigg|\frac{v}{8\pi\mu(4-\ln2)}\bigg|\bigg)
\end{equation}$$-\epsilon\frac{(213\mu-20)}{10}\bigg[\frac{v}{16\pi\mu(4-\ln2)}\bigg]$$
where $\epsilon=+1(-1)$ for AdS(dS) sector of the background space
time. By looking at the positivity condition of the specific
volumes (\ref{vA}) and (\ref{vd}) one can infer that for AdS
sector in which $x<0$ we should choose negative magnetic charge
but for dS sector in which $x>0$ we should choose $Q>0.$ In order
to plot $p-v$ diagrams at constant temperature, the critical
points are needed to be obtained from the critical equations of
$\frac{\partial t}{\partial v}|_p=0$ and $\frac{\partial^2
t}{\partial v^2}|_p=0$ which by substituting the equation of state
(\ref{ttt}) read
\begin{equation}v_c=8\pi\mu (4-\ln2)e^{-\frac{29}{6}},~~~p_c=\frac{\epsilon[45e^{-\frac{29}{3}}+426\mu-40]}{320\pi\mu(4-\ln2)},~~~t_c=\frac{\epsilon\mu e^{-\frac{29}{2}
}}{2}.\end{equation} It is useful to rewrite the equation of state
(\ref{ttt}) versus the following re-scaled thermodynamic
variables.
\begin{equation}\bar{v}=\frac{v}{v_c},~~~\bar{p}=\frac{p}{p_c},~~~\bar{t}=\frac{t}{t_c}\end{equation}
such that
\begin{equation}\label{ttbar}dS/AdS:~~~\bar{t}=a(\mu)\bar{p}\bar{v}+\frac{3}{2}\bar{v}^3\bigg[\ln \bar{v}-\frac{5}{6}\bigg]+b(\mu)\bar{v}
\end{equation}
where we defined
\begin{equation}a(\mu)=\frac{p_cv_c}{t_c}=\frac{45\mu+(426\mu-40)e^\frac{29}{3}}{20\mu},~~~b(\mu)=\frac{(20-213\mu)e^{\frac{29}{3}}}{10\mu}.\end{equation}
This equation of state has same form for both dS and AdS sector of
the space time and we plot $p-v$ diagram at constant temperature
for critical temperature $\bar{t}_c=1$ and its below values
$\bar{t}<\bar{t}_c$ and its upper values $\bar{t}>\bar{t}_c$ in
figure 2 for different values of the GB parameter. To do so we
define a critical value for the GB parameter as
$\mu_c=0.09323992798$ which is calculated by substituting
$\bar{v}=\bar{v}_c=1$, $\bar{p}=\bar{p}_c=1$ and
$\bar{t}=\bar{t}_c=1$ into the equation of state (\ref{ttbar}) and
we plot several p-v diagrams at constant temperature for
$\mu<\mu_c,$ $\mu=\mu_c$ and $\mu>\mu_c.$ These diagrams show that
 a large to small black hole phase transition for $\mu\geq\mu_c$ (see figures 1-a,b,d,f)and  small to large phase transition for
 $\mu<\mu_c$ (see figures 1-b,c,e). In the figures 1-a,b,d,f the
 diagrams have a local maximum point at higher specific volume and
 a local minimum point at smaller specific volume. It is known
 that thermodynamic stability is happened in the minimum point of
 the p-v diagram while its instability is happened at maximum
 point  of the diagram. By looking at the diagrams given at the
 figure 1 one can infer that position of the minimum and the
 maximum points of the diagrams are exchanged for $\mu\geq\mu_{c}$
 and $\mu<\mu_c$ respectively.
 In the next section
we investigate Joule-Thomson expansion of this kind of black hole
by plotting isenthalpic $t-p$ diagrams and inversion curves.
\subsection{JT Expansion}
To study the Joule Thomson adiabatic expansion of the 4D GB dS/AdS
Bardeen black hole we should investigate isenthalpic curves for
all scales of the black holes $x<1$ and $x>1$ and so we can not
use series solutions of the horizon equation and the Hawking
temperature given in the previous section for small values $x<1$
and so we must be use exact forms of these equations but we can
still use series expansions of them for small GB parameter
$\mu<1.$ Hence we first solve the horizon equation vs the
cosmological parameter $\lambda$ leading to two different solution
which one of them reaches to some physical situations where the JT
expansion is happened. By substituting it into the temperature
equation we can obtain a suitable form for the temperature
equation to be independent of the $\lambda$ term but having a long
length form. Hence we calculate its Taylor series expansion about
small values of $\mu$  which up to the second order term reads
\begin{equation}\label{lJ}
\lambda\simeq\frac{6m}{(x^2+1)^{\frac{3}{2}}}-\frac{3}{x^{2}}-\frac{3\mu}{x^5(x^2+1)^3}\times
\end{equation}
$$\bigg[4mx(x^2+1)^{3/2}(12mx\arcsin h x+5x^2+6)$$
$$-3m(x^2+1)^3(8\arcsin h x+mx^2\arctan x)-5x
(x^2+1)^3-3m^2(x^7+\frac{32}{3}x^5+15x^3)\bigg]
$$
in which $p=-\frac{\epsilon\lambda}{8\pi}$ with $(\epsilon=+(-))$
for AdS(dS) sector of the space time, is pressure of dS/AdS
background space. Moreover we approximate the relation of
temperature for small $\mu$ as follow:
\begin{equation}\label{tJ}
t\simeq \bigg(\frac{1}{x}+\lambda
x-\frac{6mx}{(x^2+1)^\frac{5}{2}}\bigg)-\mu\bigg\{\frac{3}{x^3}-\end{equation}$$\bigg[6m\bigg[\arctan
h\bigg(\frac{1}{\sqrt{x^2+1}}\bigg)-\frac{1}{\sqrt{x^2+1}}\bigg]+\frac{6}{x^2}+\lambda\ln
x\bigg]\bigg(\frac{1}{x}+\lambda
x-\frac{6mx}{(x^2+1)^\frac{5}{2}}\bigg)\bigg\}.
$$
We plot the isoenthalpic t-p curves for all scales of the 4D GB
dS/AdS Bardeen black holes with three chosen numeric values for
dimensionless GB parameter as $\mu=0.0001$, $0.001$, and $0.01$.
To determine inversion curves where the cooling phase and heating
phase of the black hole is separated with intersection point of
the inversion curve and isenthapic curves (see figures 2) we
should calculate the JT coefficient through
$\mu_{JT}=\frac{\partial t}{\partial p}|_m$ (see appendix) and
solve $\mu_{JT}=0$ for which we obtain a long length relation for
the inversion enthalpy (the mass $m_i$) and so we do not mention
it in this paper. To plot inversion curves we substitute the
inversion mass $m_i$ inside the relation of \eqref{lJ} and
\eqref{tJ} to obtain suitable relations for the inverse
temperature and inversion pressure respectively vs $x.$ Then all
ishenthalpic curves together with inversion curves are plotted for
different values of the horizon parameters $x$. These diagrams are
collected in the figure 2. One can see that left side of
intersection point of the inversion curve with the ishenthalpic
curves where the slope of the curves or the JT coefficient take on
some positive values the black hole system participate in the
heating phase but for right side of the intersection point it
participates in the cooling phase. Also one can look at the
figures 2-a,b,c,d,e to infer that by rasing the GB parameter the
black hole system at constant enthalpy take on other subsystem
which does not participate in the JT adiabatic expansion (please
compare 2-a,b,c with 2-d,e). These are shown with straight-lines
Particularly for large values of the GB parameter (see figure 2-f)
there is not a intersection between the inversion curve (solid
blue line) with the ishenthalpic curves. In fact to clear a JT
adiabatic expansion in a thermodynamic system the inversion curves
should intersect maximum point of the isenthalpic curves which are
appeared in all figures 2-a,b,c,d,e but not in the 2-f.
Furthermore we should point that the inversion curves have two
different branches which one of them intersects with the
ishenthalpic curves for small values mass of the black holes (see
figures 2-a,c,e) while for massive black hole these two branches
are approached to each other and in fact two branches of the
inversion curves intersect maximum point of the ishenthalpic t-p
curves (see figures 2-b,d).

\section{Conclusion} In this work we used a consistent model of Einstein Gauss Bonnet gravity to study thermodynamics of 4D $dS/AdS$ GB Bardeen
black hole. Metric source of this type of black hole is nonlinear
electromagnetic fields with a non-vanishing magnetic charge.
Physical importance of this type of black holes is nonsingular
property which have and they are applicable to study black hole
structure of center of galaxies. We obtain analytical solution of
the metric field equation for small values of the Gauss Bonnet
coupling constant. We calculated exterior horizon of this kind of
black holes and the Hawking temperature. At last by calculating
the equation of state in presence of both dS and AdS space
pressures we investigated the black hole thermodynamics by
focusing on the phase transitions and Joule Thomson adiabatic
expansion of the black hole in presence of Gauss Bonnet
counterterms. In this way we saw that the Gauss Bonnet coupling
constant plays important role in these phase transitions.
Positivity condition on the thermodynamic volume causes to be
valid negative (positive) magnetic charge for AdS (dS) space
pressure. Our mathematical calculations predict that 4D AdS GB
Bardeen black hole takes on small to large phase transition form
small values of the Gauss Bonnet coupling constant and vice versa
for larger values of this coupling constant. Also we understand
that for small values of the Gauss bonnet term effect the black
hole has a one subsystem which participates in the JT adiabatic
expansion but for larger values of the Gauss Bonnet effect there
is a second subsystem for the black hole which does not
participate in the cooling -Heating (JT expansion) phase
transition. As an extension of this work one can investigate JT
expansion for other types of nonsingular magnetic charged black
holes (see for instance \cite{57,58}). As an future work we like
to investigate other thermodynamic properties of the 4D GB dS/AdS
Bardeen black hole such that holographic entanglement entropy and
complexity growth rate via complexity=action conjecture.

\vskip.5cm\textbf{Appendix}\vskip .5cm As we mentioned previously
the JT expansion occurs at constant enthalpy $H=U+PV$ for which we
can write \begin{equation}dH=TdS+VdP\label{H}\end{equation} where
we used \begin{equation}TdS=dQ=dU+PdV.\end{equation} Applying
$dH=0$ the equation (\ref{H}) reduces to the form  $0=TdS+VdP$
which can be rewritten as
\begin{equation}\frac{dH}{dP}=0=T\bigg(\frac{\partial S}{\partial P} \bigg)_H+V.\label{Ts}\end{equation} If we assume that
the entropy depends to the temperature $T$ and the pressure $P$ as
$S=S(T,P)$ then we can write $dS=\big(\frac{\partial S}{\partial
P}\big)_TdP+\big(\frac{\partial S}{\partial T}\big)_PdT$ for which
we will have \begin{equation}\bigg(\frac{\partial S}{\partial
P}\bigg)_H=\bigg(\frac{\partial S}{\partial
P}\bigg)_T+\bigg(\frac{\partial S}{\partial T}
\bigg)_P\bigg(\frac{\partial T}{\partial
P}\bigg)_H.\label{sp}\end{equation} Substituting (\ref{sp}) into
the equation (\ref{Ts}) we obtain
\begin{equation}0=-T\bigg(\frac{\partial V}{\partial T}\bigg)_P+C_P\bigg(\frac{\partial
T}{\partial P}\bigg)_H+V\end{equation} where we used
$C_P=T\big(\frac{\partial S}{\partial T}\big)_P$ and the Maxwell
relationship $\big(\frac{\partial S}{\partial
P}\big)_T=-\big(\frac{\partial V}{\partial T}\big)_P$ which by
solving  $\big(\frac{\partial T}{\partial P}\big)_H$ one can
obtain the JT coefficient $\mu_{JT}=\big(\frac{\partial
T}{\partial P}\big)_H$.

\begin{figure}[!ht]
\centering  \subfigure[{}]{\label{7}
\includegraphics[width=0.34\textwidth]{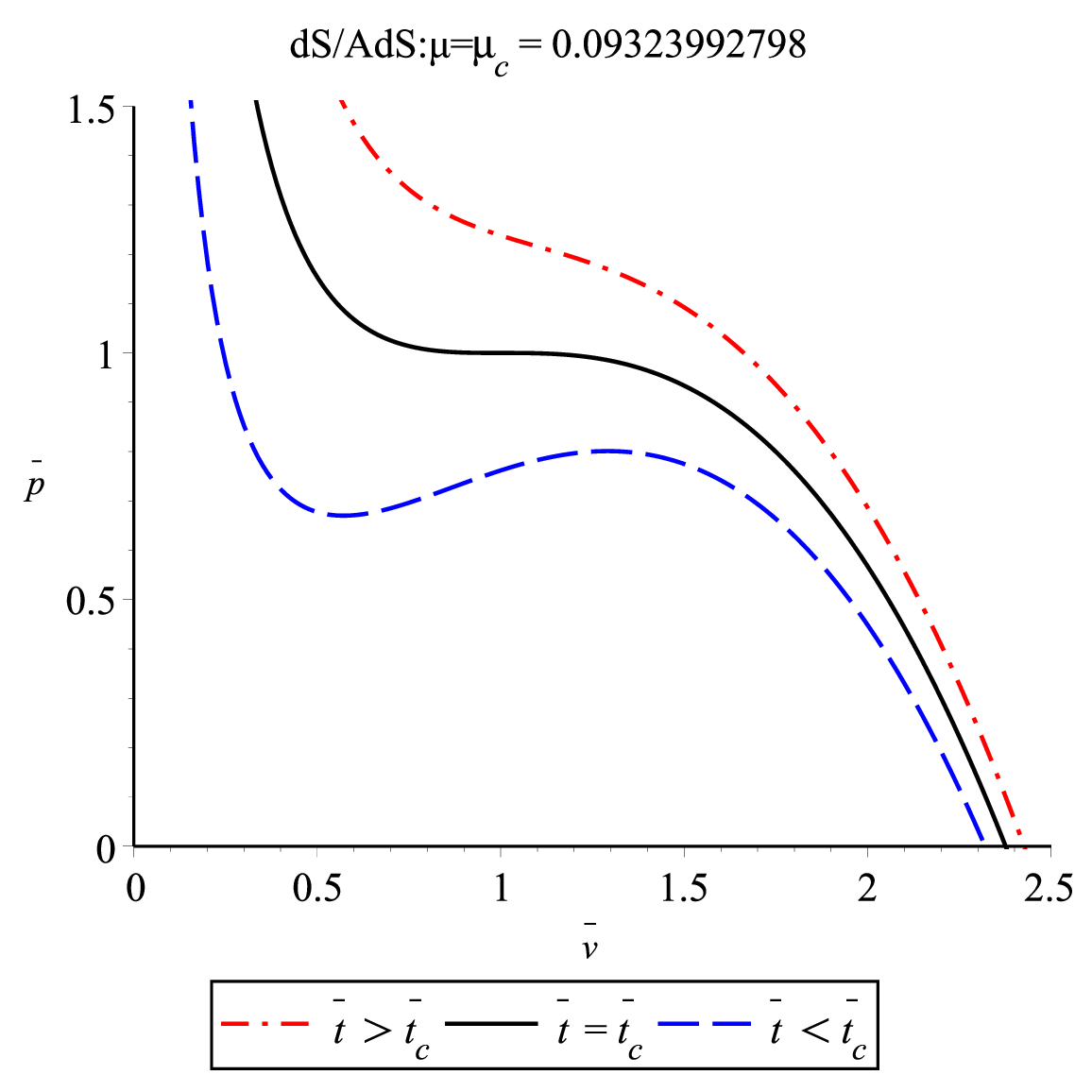}}
\centering  \subfigure[{}]{\label{8}
\includegraphics[width=0.34\textwidth]{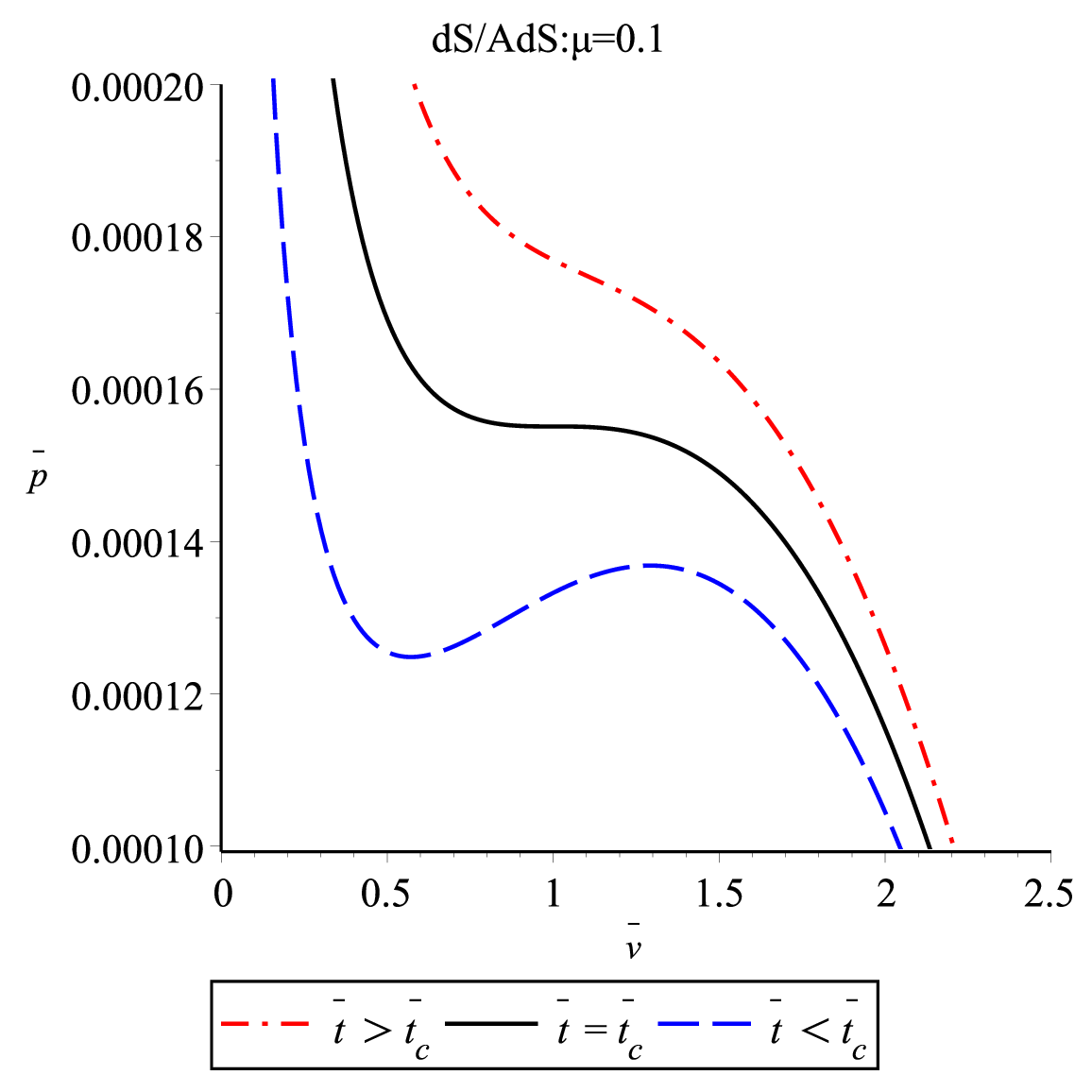}}
 \centering
\subfigure[{}]{\label{9}
\includegraphics[width=0.34\textwidth]{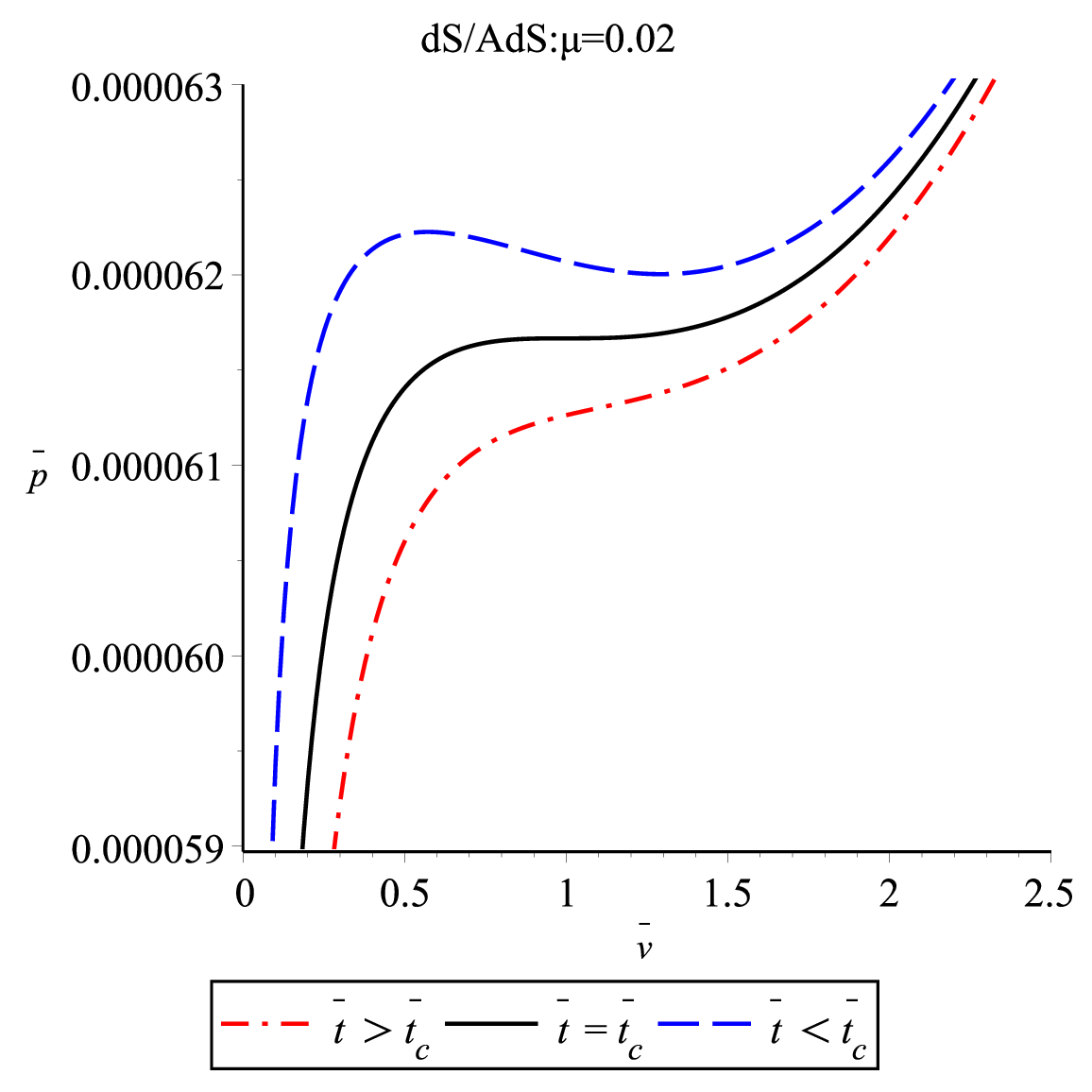}}
\centering  \subfigure[{}]{\label{10}
\includegraphics[width=0.34\textwidth]{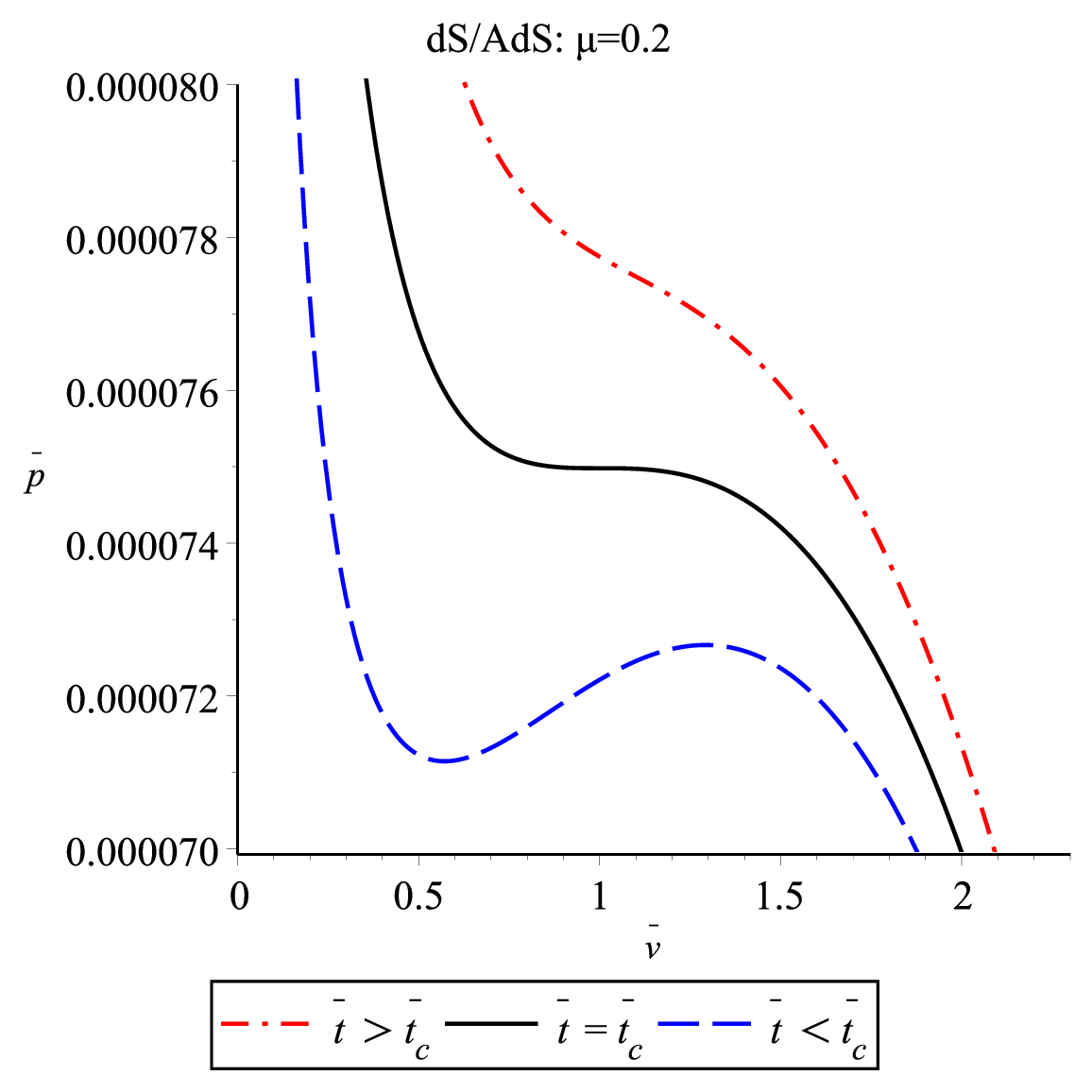}}
\centering  \subfigure[{}]{\label{11}
\includegraphics[width=0.34\textwidth]{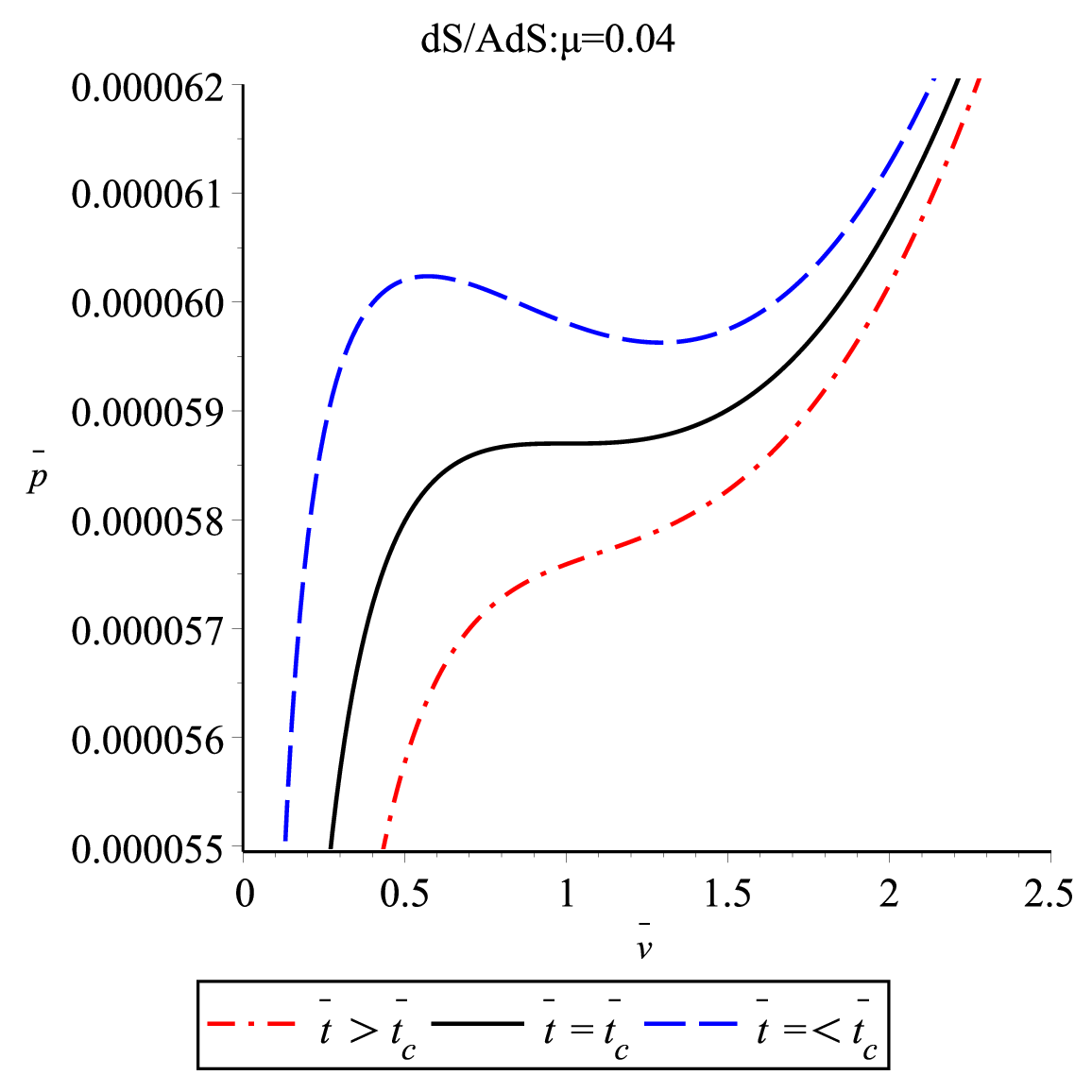}}
\centering  \subfigure[{}]{\label{6}
\includegraphics[width=0.34\textwidth]{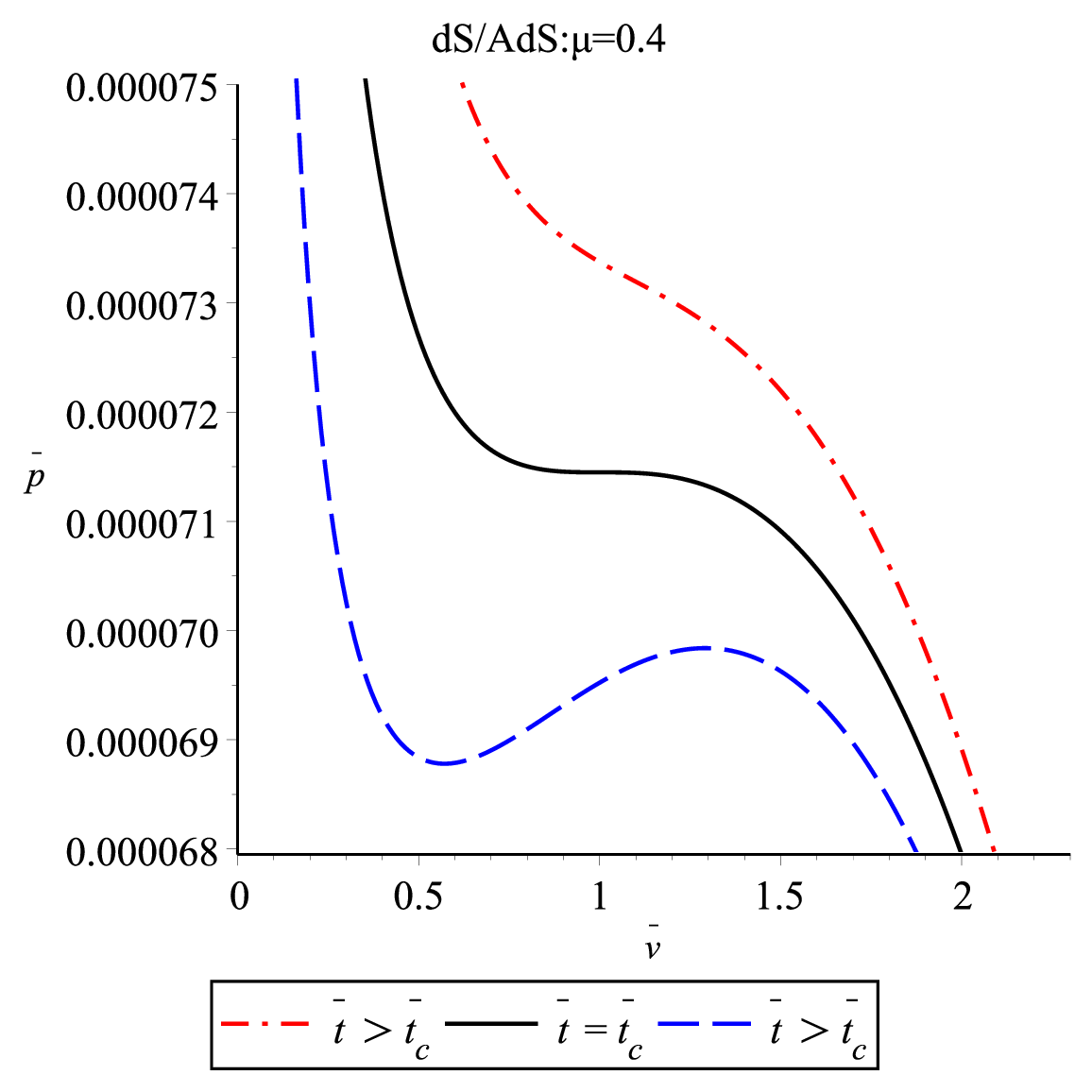}}
\hspace{3mm} \caption{\footnotesize{p-v diagrams for small scale
4D GB dS/AdS Bardeen black holes}}
\end{figure}

\begin{figure}[!ht]
\centering  \subfigure[{}]{\label{101}
\includegraphics[width=0.34\textwidth]{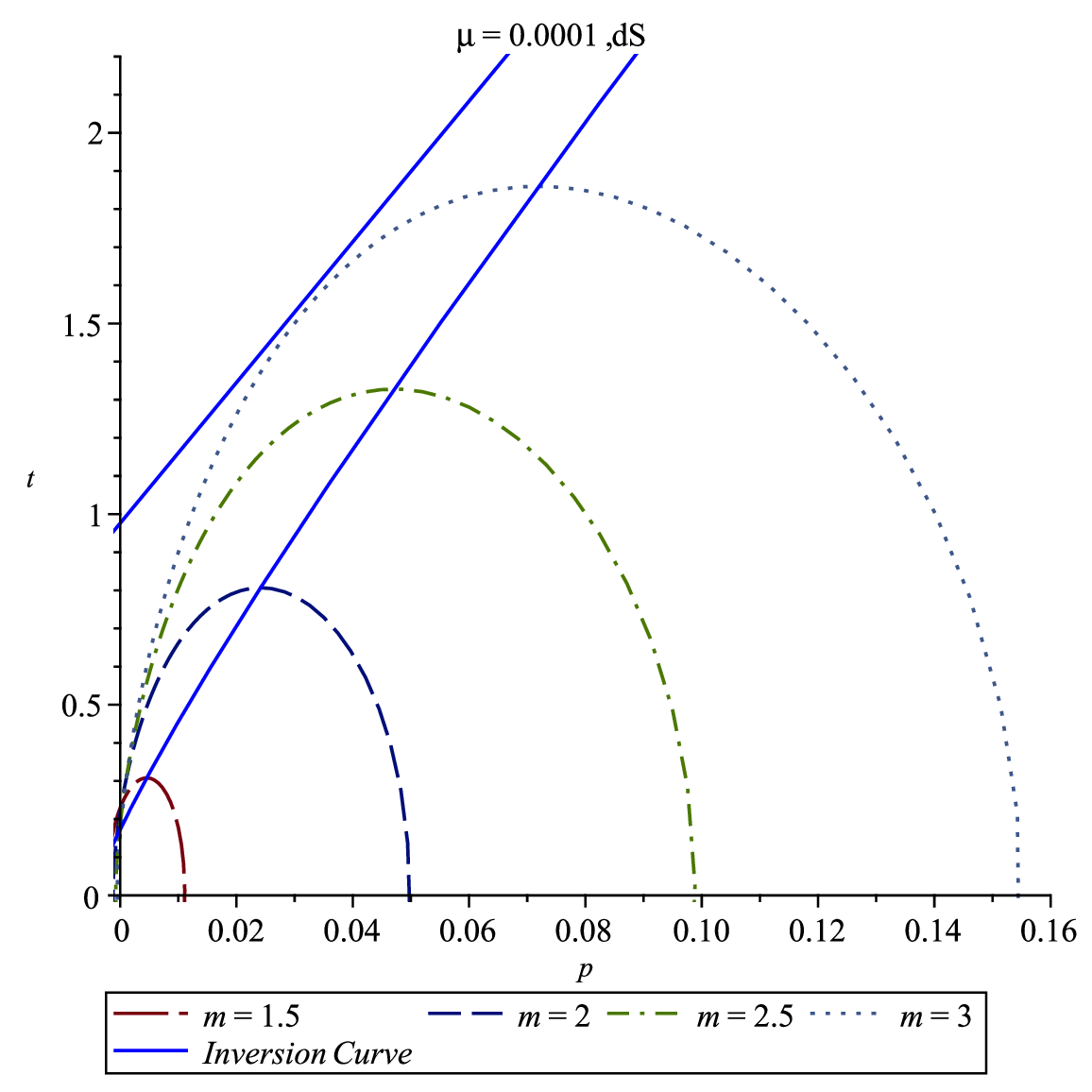}}
\centering  \subfigure[{}]{\label{102}
\includegraphics[width=0.34\textwidth]{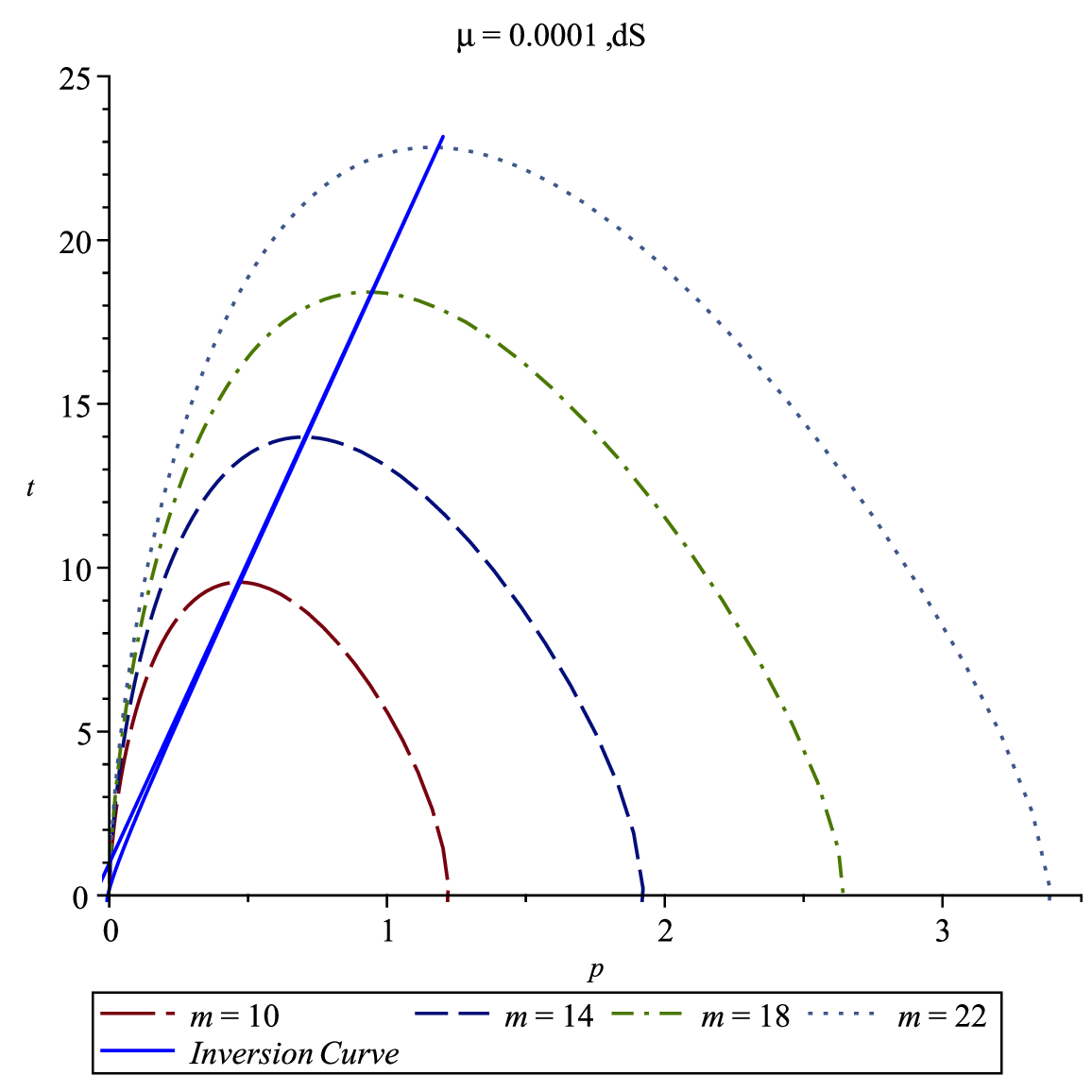}}
 \centering
\subfigure[{}]{\label{103}
\includegraphics[width=0.34\textwidth]{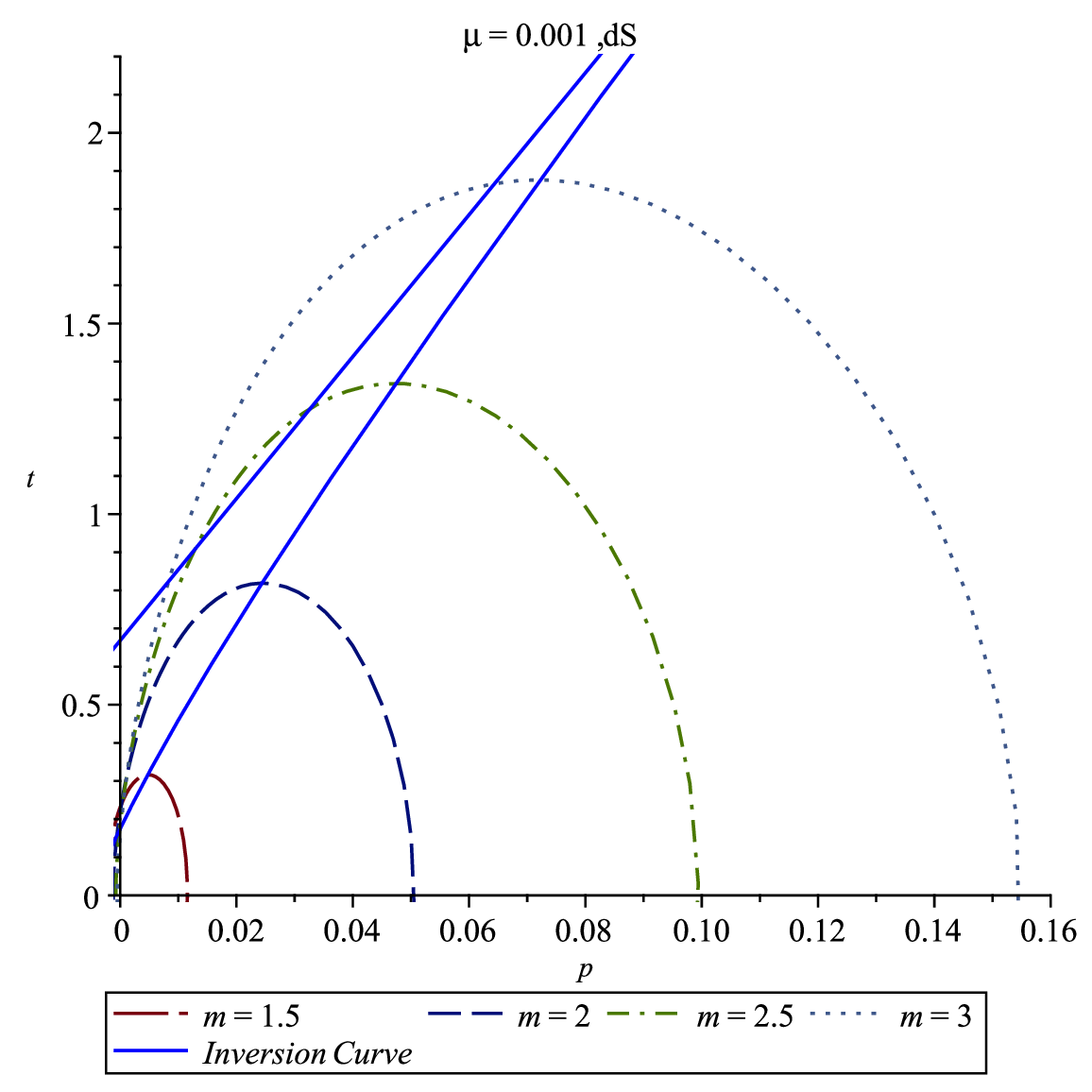}}
\centering  \subfigure[{}]{\label{104}
\includegraphics[width=0.34\textwidth]{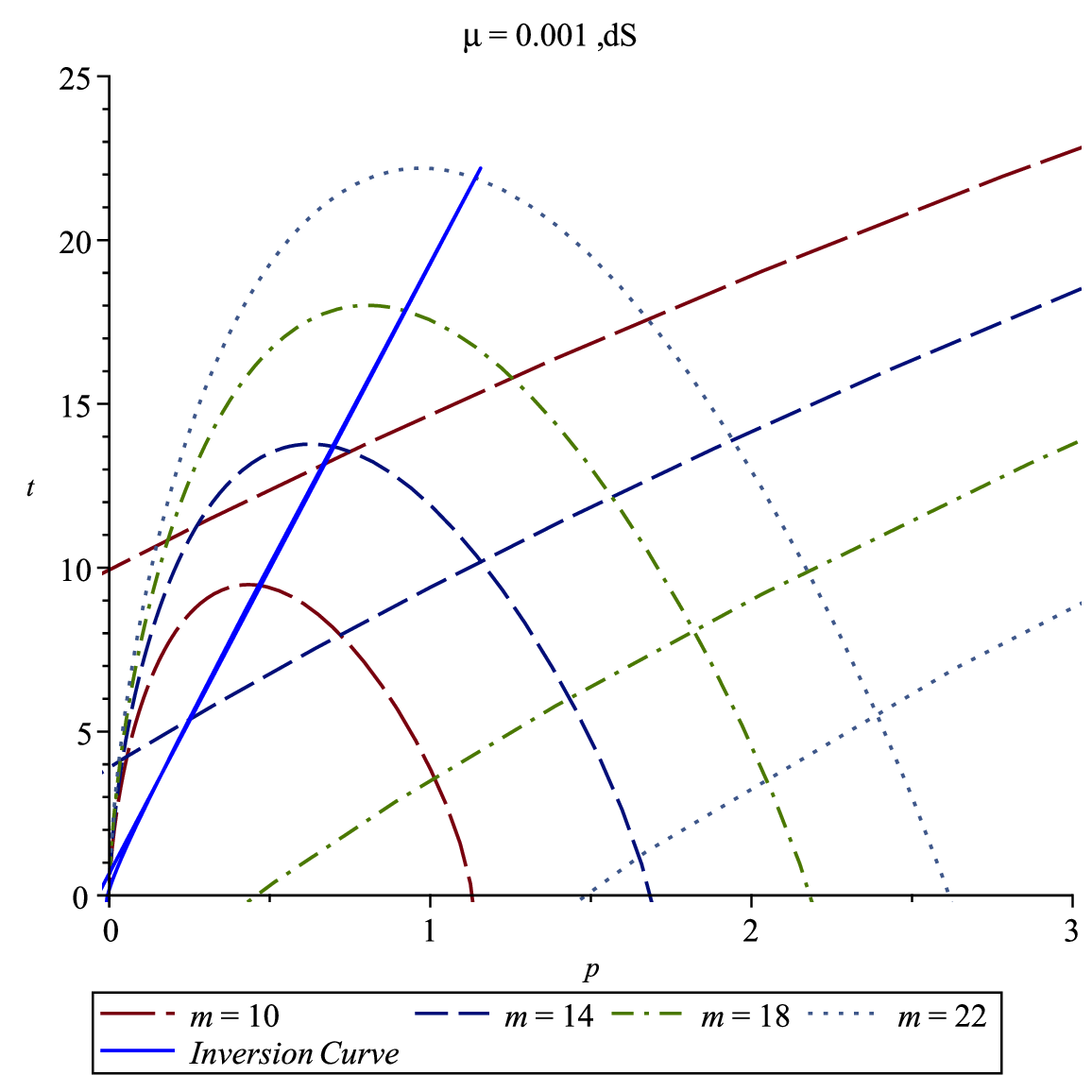}}
\centering  \subfigure[{}]{\label{105}
\includegraphics[width=0.34\textwidth]{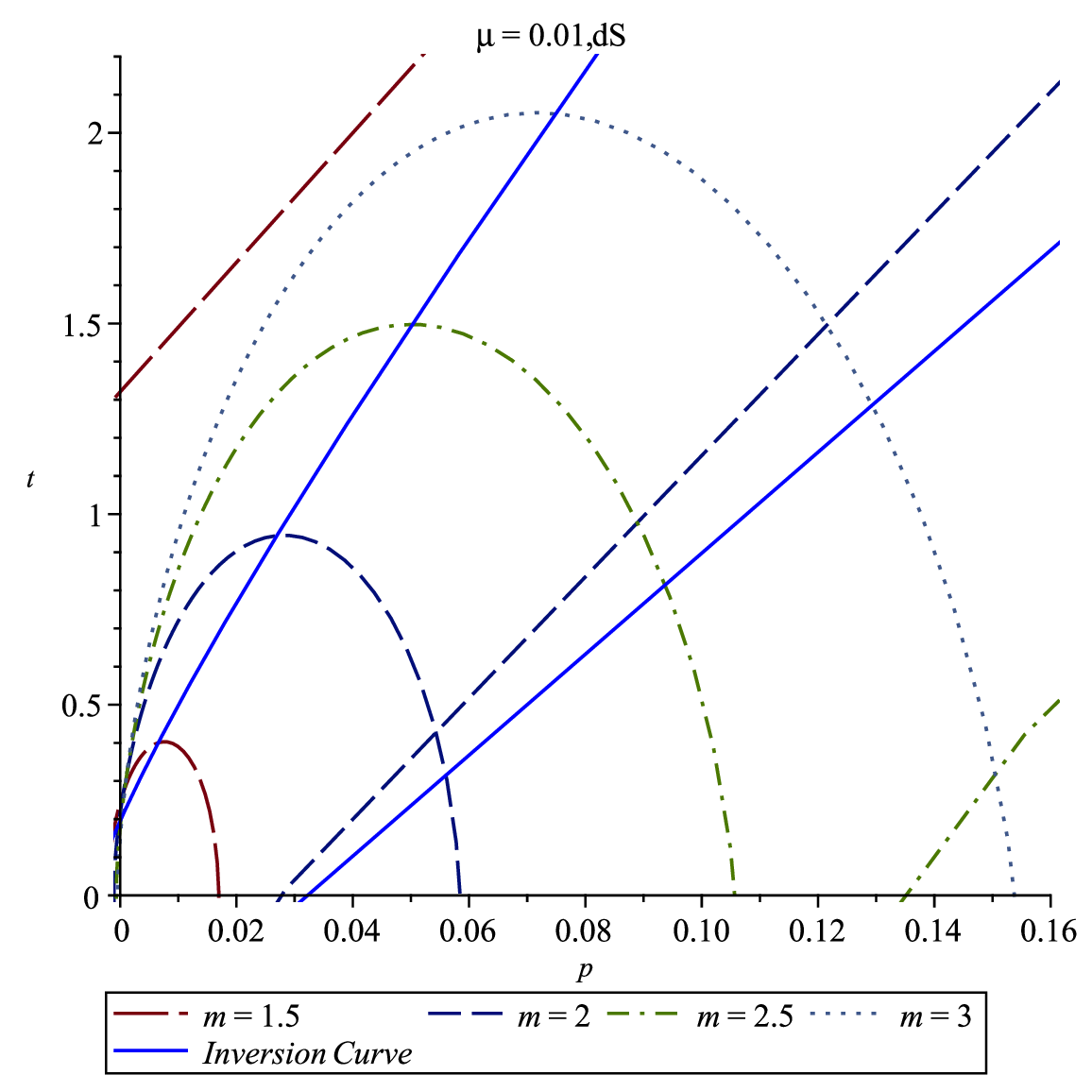}}
\centering  \subfigure[{}]{\label{106}
\includegraphics[width=0.34\textwidth]{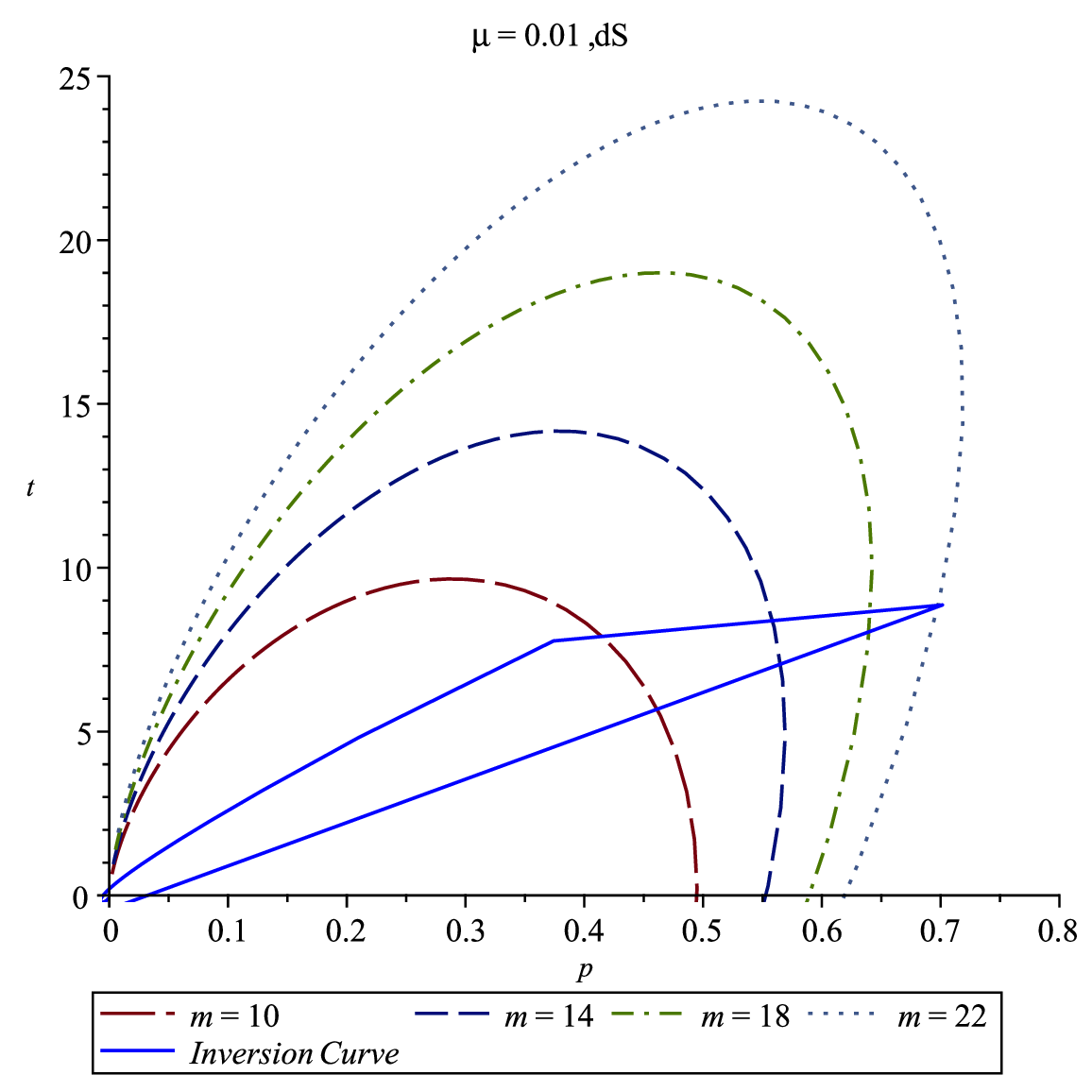}}
\hspace{3mm} \caption{\footnotesize{Isenthalpic and inversion
curves for 4D GB dS/AdS Bardeen black holes with different values
for the enthalpy $m$}}
\end{figure}

\end{document}